\documentclass[twocolumn,showpacs,preprintnumbers,amsmath,amssymb,letterpaper]{revtex4}

\usepackage{graphicx}
\usepackage{amsmath}
\usepackage{epsfig}

\begin{document}
%
%
%
%
%
\def\bra#1{\mathinner{\langle{#1}|}}
\def\ket#1{\mathinner{|{#1}\rangle}}
\def\braket#1{\mathinner{\langle{#1}\rangle}}
\def\Bra#1{\left<#1\right|}
\def\Ket#1{\left|#1\right>}
{\catcode`\|=\active
  \gdef\Braket#1{\left<\mathcode`\|"8000\let|\BraVert {#1}\right>}}
\def\BraVert{\egroup\,\mid@vertical\,\bgroup}
%

{\catcode`\|=\active
  \gdef\set#1{\mathinner{\lbrace\,{\mathcode`\|"8000\let|\midvert #1}\,\rbrace}}
  \gdef\Set#1{\left\{\:{\mathcode`\|"8000\let|\SetVert #1}\:\right\}}}
\def\midvert{\egroup\mid\bgroup}
\def\SetVert{\egroup\;\mid@vertical\;\bgroup}

%

\title{Renormalized mean-field theory for a two-component Fermi gas with s-wave
interactions.}
\author{Javier von Stecher}
\author{Chris H. Greene}
\affiliation{Department of Physics and JILA, University of Colorado,
Boulder, Colorado 80309-0440}

\begin{abstract}
A method is introduced to renormalize the zero-range interaction for
use in mean-field and many-body theory, starting from two-body
calculations. The density-renormalized delta-function interaction is
then applied using mean-field theory to a two-component fermion gas,
and compared with diffusion Monte Carlo simulations and conventional
mean-field calculations. In the unitarity limit, the equation of
state exhibits the expected behavior $\mu\propto\rho^{2/3}$, with a
parameter $\beta= -0.492$, which is consistent with recent
experiments\cite{partridge2006pap,bourdel2004esb,kinast2005hcs,Stewart06}.
\end{abstract}
\maketitle

\section{Introduction}

The low energy scattering of fermionic atoms controls the structure
and dynamics of an ultracold quantum Fermi gas. When the scattering
length $a_0$ between fermions in different internal spin states is
tunable, for instance, in an external magnetic field, it becomes
possible to study the crossover between BCS-type superfluidity of
momentum pairs and Bose-Einstein condensation (BEC) of molecules.

In recent years, the BCS-BEC crossover problem has become
experimentally accessible
\cite{greiner2003emb,jochim2003bec,zwierlein2004cpf,kinast2004esr},
enabling sharp tests. The BCS theory has been successful in
explaining superfluidity in Fermi gases, but this theory is
incomplete because it neglects the Hartree term of the interaction,
$4\pi a \rho/m$, where $\rho$ is the density of one spin component.
Comparatively little research has considered Fermi gases including
the Hartree term, with the primary regime studied being the
perturbative case where $\rho a_0^3<<1$. This is usually referred as
the ``normal state'' of the gas, in contrast with the superfluid
state. Quantum Monte Carlo (QMC) simulations include both the
Hartree term and pairing physics, but a complete theory that
contains both ingredients is still required.

When the range of interaction is much smaller than the
inter-particle distance, the potential can be replaced by a delta
function interaction although this must be done with caution because
the delta function interaction is too singular to be exactly
solvable, even in principle. In the weakly-interacting limit, $\rho
a_0^3<<1$, the coupling parameter in the delta function interaction
is proportional to the two-body scattering length $a_0$; this is
known as the Fermi pseudo-potential \cite{fermi36}. Using this
approximation, mean field theories have been applied to Fermi gases
\cite{bruun1998ifg,Rittenhousecondmat}. The use of this
approximation in strong interacting regimes leads to an unphysical
collapse of the Fermi gas. To go beyond the Fermi approximation for
the pseudo-potential, it is crucial to renormalize the coupling
constant; the purpose of this paper is to introduce a new and
convenient way to achieve this renormalization.

Full diagonalization of a Hamiltonian with delta-function
interactions requires a momentum cut-off renormalization even in the
weak interacting limit. This type of renormalization has been
carried out, for example, in
Refs.\cite{KokkelmansHolland,BraatenNieto,lepage9706029rse,romans2006btf,javanainen2005smf,HollandLevin,sheehy2006bbc,ohashi2002bbc,kokkelmans2004dam},
just to name a few such studies. However, such a renormalization is
unnecessary at small or modest scattering lengths, when treated by
mean-field theories with zero-range interactions because they are
well behaved in this limit. To go beyond the weakly-interacting
limit of mean-field theory we propose a density-dependent
renormalization of the coupling parameter that is intended to apply
even in the long-wavelength limit. A density-dependent
renormalization for a 2-component degenerate Fermi gas has been
recently proposed in Ref. \cite{Baym} to explain the stability of
this system in the strong interacting limit; a functional form for
the effective scattering length $a_{eff}$ was designed to give the
expected behavior (i.e. as determined by QMC and other calculations)
in both the weak interaction limit and the unitarity limit.  Our
renormalization strategy is different, in that we present a method
to calculate $a_{eff}$ by using the exact energies of two particles
in a trap. We compare the many-particle predictions obtained using
the renormalized interaction potential with diffusion Monte Carlo
and alternative mean-field calculations, and find that our
renormalization automatically gives the correct behavior in both the
strong and weak interaction limits, for both positive and negative
scattering lengths, without imposing this constraint at the outset.

This paper is organized as follows. In Section II we develop the
renormalization procedure and show that a simple 2-parameter
analytical formula can be utilized to an excellent approximation
over the whole range from positive to negative two-body scattering
lengths. Section III applies the renormalization to many-particle
mean-field theory and presents some of its predictions. Section IV
compares the results obtained using our renormalized mean-field
theory with quantum Monte Carlo calculations and with perturbative
mean-field calculations.

\section{Renormalization procedure}

Through this paper we consider a system of equal mass fermions in a
spherically symmetric harmonic oscillator trap at temperature $T=0$.
The Hamiltonian that we adopt is

\begin{equation}
\label{Ham} \mathcal{H}=\sum_i \left(
-\frac{\hbar^2}{2m}\nabla^2_i+\frac{1}{2}m\omega^2\mathbf{r}^2_i\right)+
\sum_{i<i'}\frac{4\pi\hbar^2a_{eff}}{m}
\delta(\mathbf{r}_i-\mathbf{r}_{i'})
\end{equation}

This Hamiltonian cannot be diagonalized exactly, since the delta
function interaction is too singular and  would produce divergent
results.\cite{esry1999vsi} The level of approximation we adopt to
diagonalize the many-body Hamiltonian is the same we use to solve
the corresponding Schr\"odinger equation for the two-body system.
For example, if we want to diagonalize this Hamiltonian in the RPA
approximation, we would use RPA for the two-body system and obtain
the renormalization through the matching procedure explained later
in this Section.  An explicitly correlated wavefunction or an
extensive configuration interaction (CI) wavefunction can produce
divergent results and a momentum cutoff renormalization is
necessary. Since we want to obtain and apply a density
renormalization without the necessity of introducing a momentum
renormalization, we need to carefully select the level of
approximation in the wavefunction.  The Hartree-Fock (HF)
approximation does not introduce explicit interparticle
correlations, as the only correlations included are ``exchange
correlations'' from the Pauli exclusion principle. This makes the HF
wavefunction a suitable approximation to adopt in our
renormalization technique, since it does not require a momentum
renormalization.

To obtain the renormalized scattering length we solve (\ref{Ham})
for two opposite-spin fermions in the HF approximation. The ground
state energies of this approximation are matched with the exact
energies of the system for different values of the bare two-body
scattering length $a_0$. From this procedure we extract the
functional dependence of $a_{eff}$ on $a_0$. The spectrum of two
opposite-spin fermions in a trap having a specified scattering
length $a_0$ and zero-range interactions can be determined
exactly\cite{busch1998tca,blume2002fpa,borca2003tap,bolda2002esl} and
(\ref{Ham}) can be solved numerically for two particles using a
Hartree-Fock wavefunction.

Dimensional analysis suggests that, in an infinite, uniform Fermi
gas, where the range of the two-body interaction is much smaller
than both the average interparticle distance and the bare scattering
length $a_0$, the only parameter that characterizes the behavior of
the system is the dimensionless combination $k_fa_0$ of the Fermi
momentum and $a_0$. Throughout this paper the Fermi momentum is
defined as $k_f\equiv (6\pi^2\rho)^{1/3}$ where the density is just
the one-spin component density. If we were applying the renormalized
scattering length $a_{eff}$ to an infinite uniform system, the only
relevant parameter would be $k_fa_{eff}$. This suggests that
$k_fa_{eff}$ has to be a function of $k_fa_0$. So, we propose the
following functional dependence,

 \begin{equation}
  \label{aeff}
    a_{eff}\equiv\frac{\zeta(k_fa_0)}{k_f}
\end{equation}

We will see below that the renormalization function,
$\zeta(k_fa_0)$, will have the desired behavior in the limiting
cases, becoming independent of $k_fa_0$ in the unitarity limit
($|a_0|\rightarrow\infty$) and reproducing the relation
$a_{eff}=a_0$ in the weak interacting limit ($k_fa_0\ll1$). We
consider that Eq.(\ref{aeff}) holds even with the inclusion of a
trapping potential. The renormalized scattering length $a_{eff}$ can
be viewed as accounting, in some way, for the correlations neglected
in the mean-field wavefunctions.

In choosing to extract $a_{eff}$ from a two-body system, we are
implicitly assuming that two-body correlations are the most
important in the many-body system. This assumption is reasonable for
two-spin component fermions with short-range interactions because
the probability of finding more than two fermions close enough to
interact is usually negligible.

\subsection{Exact Energies}

This subsection reviews the exact results for two particles
in a trap interacting through a zero-range pseudopotential. Then, we will select
from the energy spectrum, the energy branch that is of interest to study in the
renormalization procedure.

Consider two particles of mass $m$ interacting through a two-body
potential $V(\mathbf{r})$ in a spherically symmetric trap. If the
effective range of the two-body potential is much smaller than the
caracteristic length of the trap, the low-lying energy levels depend only on the
scattering properties of the potential and not on its shape. Under
this condition, the two-body potential can be replaced by a
pseudopotential of the form \cite{HuangYang}:
 \begin{equation}
  \label{pseudopot}
   v(\mathbf{r})=\frac{4\pi\hbar^2a(E)}{m} \delta(\mathbf{r})\frac{\partial}{\partial r}r,
\end{equation}
where $a(E)$ is the energy-dependent scattering length. For
ultra-cold gases, the energy dependence on the scattering length can
be neglected, so $a(E)$ can be replaced by its energy-independent limit $a_0$.

Two particles in a trap with this pseudopotential have been
considered previously
\cite{busch1998tca,blume2002fpa,borca2003tap,bolda2002esl}. After
separating the center of mass and relative coordinates, the problem
reduces to two independent one-dimensional Schrodinger equations. The
pseudopotential (\ref{pseudopot}) is introduced as a boundary
condition in the relative coordinate Schrodinger equation. The
energies are
\begin{equation}
  \label{ExE}
  E_{exact}=E_{CM}+E_{rel},
  \end{equation}
 where $E_{CM}=(n_{CM}+3/2)\hbar\omega$ and
\begin{equation}
  \label{ExE2}
  \sqrt{2}\frac{\Gamma\left(-\frac{E_{rel}}{2\hbar\omega}+
  \frac{3}{4}\right)}{\Gamma\left(-\frac{E_{rel}}{2\hbar\omega}+
  \frac{1}{4}\right)}=\frac{a_{ho}}{a_0}
\end{equation}

\begin{figure}[h]
\includegraphics[scale=0.6]{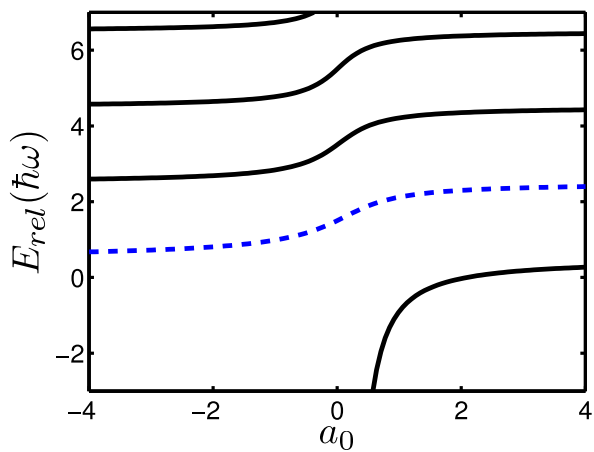}
\caption{ (Color Online) Spectrum $E_{rel}$ as a function of $a_0$.
The dashed blue line corresponds to the energy branch selected for
the renormalization procedure developed in this study.
\label{2pspec}}
\end{figure}

The trap length $a_{ho}$ is defined in this case as
$a_{ho}=\sqrt{\hbar/m\omega}$. Equations (\ref{ExE}) and
(\ref{ExE2}) reproduce the complete spectrum of the system with zero
relative angular momentum. Figure \ref{2pspec} shows the spectrum of
$E_{rel}$ (Eq. (\ref{ExE2})) as a function of the scattering length.
The lowest curve of the spectrum shown describes the formation of a
molecule, where interparticle correlations are fundamental.
One anticipates that a HF
wavefunction would be a terrible approximation to such a state
in which the two atoms are bound together to form a bound
molecular eigenstate. Since in this work
we do not consider molecule formation, we will
consider instead the second branch in Fig. \ref{2pspec} for the
renormalization. The renormalization for positive scattering length
will only be valid when two-body potential that does not support a
bound state. The energies used for the renormalization are
$E_{CM}=3/2\hbar\omega$ and the energy branch where
$1/2\hbar\omega<E_{rel}<5/2\hbar\omega$. This branch of solutions is
a smooth curve which gives the correct non-interacting energy at
$a_0=0$.

\subsection{Mean-field solution}

Next we determine a renormalization function
$\zeta(k_fa_0)$ which, applied to the system of two particles in a
trap and using the HF approximation, yields the exact results
obtained in the previous section.

To obtain the HF solution of two opposite-spin particles in a trap,
we utilize a product wavefunction having the same orbital for both
particles. Thus, the two-body spatial wavefunction is
\begin{equation}
  \label{wavefun}
   \Psi(\mathbf{r}_1,\mathbf{r}_2)=\psi(\mathbf{r}_1)\psi(\mathbf{r}_2)
\end{equation}
and the spin part is antisymmetric. We introduce this trial
wavefunction into the Hamiltonian and obtain the energy functional,

\begin{eqnarray}
  \label{Efun}
   \mathcal{E}(\psi)=\int \left(2\psi(\mathbf{r})
   \left(
  -\frac{\hbar^2}{2m}\nabla^2+\frac{1}{2}m\omega^2\mathbf{r}^2\right)\psi(\mathbf{r})\right.\nonumber \\
  \left.+\frac{4\pi\hbar^2 \zeta(k_fa_0)}{mk_f} \psi(\mathbf{r})^4\right) d\mathbf{r}.
\end{eqnarray}
Minimization of this energy functional determines the ground state
energy and wavefunction. The minimization is done with respect to
the orbital $\psi(\mathbf{r})$ as in a standard HF
procedure.\cite{EsryThesis} But prior to carrying out this
minimization of Eq. (\ref{Efun}), we must choose how to evaluate
$k_f$.  Since its formal definition is $k_f\equiv
(6\pi^2\rho)^{1/3}$, this means that $k_f$ depends at each $r$-value
on $\psi(\mathbf{r})$. For many-particle systems, we would use local
density approximation to evaluate $k_f$. But the application of a
local density approximation for a system of two particles does not
seem physically correct, so, for two particles we consider the
expectation value of $\overline{k}_f$, to be the more appropriate
quantity.
\begin{equation}
  \label{kfave}
\overline{k}_f\equiv\int k_f(\mathbf{r}) \psi(\mathbf{r})^2
d\mathbf{r}= \int (6 \pi^2\psi(\mathbf{r})^2)^{1/3}
\psi(\mathbf{r})^2 d\mathbf{r}
\end{equation}
The minimization procedure leads to a Schr\"odinger-type equation,
where $\psi^2(\mathbf{r})$ is the 1-particle density:
  \begin{multline}
  \label{GP}
 \left( -\frac{\hbar^2}{2m}\nabla^2+\frac{1}{2}m\omega^2\mathbf{r}^2+\right. \\
 \frac{8\pi\hbar^2
(6\pi^2)^{1/3}\overline{\rho}}{3m}\left(\frac{\zeta^\prime(\overline{k}_f
a_0) a_0}{\overline{k}_f}-\frac{\zeta(\overline{k}_f
a_0)}{\overline{k}_f^2}\right) \psi(\mathbf{r})^{2/3} \\
\left.+\frac{4\pi\hbar^2 \zeta(
 \overline{k}_f a_0)}{m\overline{k}_f}\psi(\mathbf{r})^2
 \right) \psi(\mathbf{r})=\epsilon\psi(\mathbf{r})
 \end{multline}
where $\epsilon$ is a Lagrange multiplier which represents the
chemical potential. The relation between $\epsilon$ and the energy
is not as straightforward as in the HF case, owing primarily to the
appearance of $\zeta^\prime(k_fa_{0})$. It should be understood that
$\psi(r)^{2/3}$ is supposed to be evaluated on a branch for which it
is real and positive everywhere.  Here and in the following,
$\zeta^\prime(x) \equiv d\zeta(x)/dx$.

Equation (\ref{GP}) corresponds to the GP equation for 2 particles
with a renormalized scattering length $
a_{eff}=\zeta(\overline{k}_fa_0)/\overline{k}_f$. After solving
Eq.(\ref{GP}) we use (\ref{Efun}) to evaluate the energy. The basic
idea is, for any chosen bare two-body scattering length $a_0$, to
find $\zeta(\overline{k}_fa_{0})$ so that the energy of the ground
state of (\ref{Efun}) matches exactly the appropriate energy of
(\ref{ExE}). From our numerical experience, the functional
dependence of $\zeta$ on $\overline{k}_f a_0$ appears to be uniquely
defined by the set of equations (\ref{ExE}, \ref{ExE2}, \ref{Efun},
\ref{kfave}) and (\ref{GP}).

There are two self-consistent procedures involved in this
calculation. To solve Eq. (\ref{GP}) we follow the standard HF
procedure, in which we adopt noninteracting solutions as the initial
guess for the orbitals, after which we iterate Eq. (\ref{GP}) until
convergence is achieved. For this procedure, we need the functional
form of $\zeta(\overline{k}_f a_0)$ and $\zeta^\prime(\overline{k}_f
a_0)$ over a range of $\overline{k}_f a_0$ values since
$\overline{k}_f$ is changing in each iteration. This means that we
cannot find the exact renormalization function $\zeta(\overline{k}_f
a_0)$ at any fixed value of $a_0$ without knowledge of the
functional form of $\zeta(\overline{k}_f a_0)$ at nearby values. To
solve this problem we calculate $\zeta(\overline{k}_f a_0)$
self-consistently over the {\it entire} range in $\overline{k}_f
a_0$ that is of interest. First we select a set of scattering length
$a_0$ values which cover the entire range of interest. For an
initial trial dependence $\zeta^{(0)}(\overline{k}_f a_0)$ we solve
Eqs. (\ref{GP}) and (\ref{Efun}) at each $a_0$, obtaining the energy
$E$, $\overline{k}_f$, the wavefunction and $\epsilon$. Then, to
obtain a new $\zeta^{(1)}(\overline{k}_f a_0)$, we look for the
value of scattering length $\widetilde{a}_0$ for which
$E_{exact}(\widetilde{a}_0)=E$, and we generate a new
renormalization function that satisfies $\zeta^{(1)}(\overline{k}_f
\widetilde{a}_0)=\zeta^{(0)}(\overline{k}_f a_0)$. The modification
of the renormalization function is evidently in the abscissa rather
than in the ordinate. This is a convenient way to approach this
calculation. Once we have carried out the matching procedure onto
the whole set of scattering length $a_0$ values, we generate the
next iteration for $\zeta^{(1)}(\overline{k}_f a_0)$ and its
derivative by interpolation.

In the next iteration, $\zeta^{(0)}(\overline{k}_f a_0)$ is replaced by
$\zeta^{(1)}(\overline{k}_f a_0)$ and we repeat the energy
matching step for the whole set of $a_0$-values. This procedure is
repeated a few times until it converges to give a single correct
renormalization function function $\zeta(\overline{k}_f a_0)$.
Note that this iterative
procedure determines a ``numerically exact'' renormalization function
$\zeta(\overline{k}_f a_0)$. Because the iteration procedure is
efficient, in 5 iterations we obtain 9 digits of agreement
between $E_{exact}$ and $E$ over the entire $a_0$ range. It is
important to introduce a sensible initial trial renormalization
function $\zeta^{(0)}(\overline{k}_f a_0)$. Many trial
$\zeta^{(0)}(\overline{k}_f a_0)$ functions, like
$\zeta^{(0)}(\overline{k}_f a_0)=\overline{k}_f a_0$, would produce
collapse of the two-fermion wavefunction for large and
negative $a_0$. To avoid
this collapse, we propose an initial trial $\zeta^{(0)}(\overline{k}_f a_0)$
which is close to the correct $\zeta(\overline{k}_f a_0)$, this is
done by choosing a qualitatively correct functional form with a few
free parameters and we then find the set of parameters that best reproduce
the exact two-body energies.

\begin{figure}[h]
\includegraphics[scale=0.6]{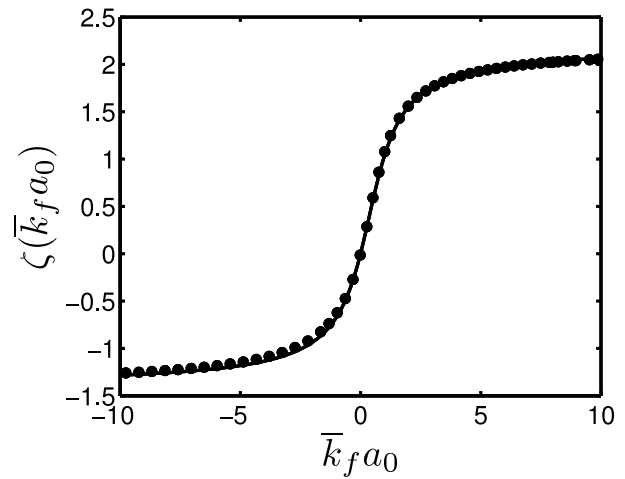}
\caption{Effective scattering length $\zeta(\overline{k}_f a_0)$
(circles) and its analytical  approximation $\zeta_0(\overline{k}_f
a_0)$ (full line) \label{zetafig}}
\end{figure}

\begin{figure}[h]
\includegraphics[scale=0.6]{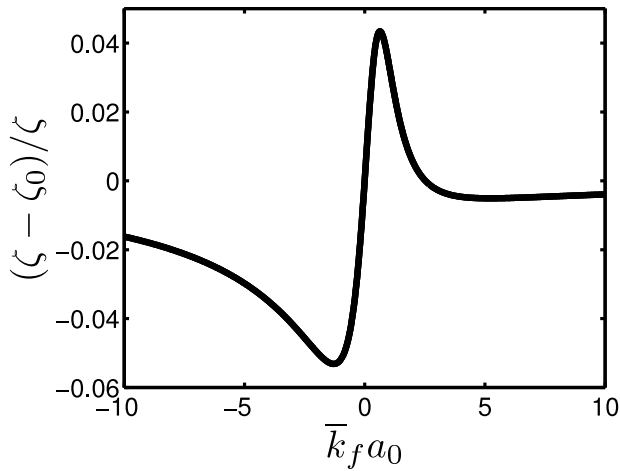}
\caption{Fractional error in our analytical approximation to the
numerical renormalization function, $\zeta_0(\overline{k}_f a_0)$.}
\label{zetaerror}
\end{figure}

The final numerical results obtained for the renormalization
function $\zeta(\overline{k}_f a_0)$ are accurately approximated by
the monotonic functional form $\zeta_0(\overline{k}_f a_0) =A+B
\arctan(C+D \overline{k}_f a_0)$, where A and B are chosen to have
the corresponding maximum and minimum values at
$a_0\rightarrow\pm\infty$ and C and D are chosen to obey
$\zeta(\overline{k}_f a_0) \rightarrow \overline{k}_f a_0$ for
$\overline{k}_f a_0 <<1$. The maximum value is $\zeta^{max}=2.182$
and value is $\zeta^{min}=-1.392$, this leads to $A=0.395$ and
$B=-1.138$. To get the correct behavior for $\overline{k}_f a_0
<<1$, this in turn requires $C\equiv\arctan(-A/B)\approx 0.362$ and
$D\equiv -(1+C^2)/B\approx -0.994$. Thus there are only two
independent parameters $A,B$ to be specified at this level of
approximation. Figure \ref{zetafig} compares our numerical results
for $\zeta(\overline{k}_f a_0)$ with this arctangent approximation,
\begin{equation}
    \label{zeta0}
\zeta_0(\overline{k}_f a_0) =0.395-1.138 \arctan(0.362-0.994
\overline{k}_f a_0).
\end{equation}
Figure \ref{zetaerror} displays the fractional error in
$\zeta_0(\overline{k}_f a_0)$ defined as $(\zeta(\overline{k}_f
a_0)-\zeta_0(\overline{k}_f a_0))/\zeta(\overline{k}_f a_0)$,
showing a maximum error of approximately $5\%$.

\begin{table}
\caption{\label{zetavalues} Exact numerical values of $\zeta(x)$.}
\begin{ruledtabular}
\begin{tabular}{cccccccc}
 x&$\zeta(x)$& x &$\zeta(x)$& x &$\zeta(x)$\\
 \hline $-\infty$& -1.392&  -1.6582& -0.82404&4.9199& 1.9248\\
 -11.937&-1.2818&   -1.3097 &-0.73716&  5.2872& 1.9426\\
 -11.394& -1.2767&  -0.96693& -0.62355&5.6545& 1.9581\\
 -10.85 &-1.2712&  -0.63317& -0.47254& 6.0219& 1.9717\\
 -10.306& -1.2651& -0.31351& -0.27173&6.3892& 1.9838\\
 -9.7626& -1.2584&  -0.014329& -0.014242& 6.7566& 1.9946\\
 -9.2193& -1.251& 0.26087& 0.2863&7.1239 &2.0042 \\
 -8.6758& -1.2428&0.51684& 0.59122&7.4913& 2.0129 \\
 -8.1326& -1.2336&0.76339& 0.86101&7.8586 & 2.0208 \\
 -7.5895& -1.2231&1.0072& 1.0787 & 8.226& 2.028\\
 -7.0467& -1.2113&1.2507& 1.2474& 8.5933& 2.0346\\
 -6.504& -1.1977&1.6166& 1.4309&8.9607 &2.0406 \\
 -5.9617& -1.182& 1.9829& 1.5585&8.0056 & 2.0238 \\
 -5.4197 &-1.1635&2.3497& 1.6508& 8.9117 &2.0398 \\
 -4.8782& -1.1416& 2.7166& 1.7201& 9.523 &2.04896\\
 -4.3373& -1.1153&3.0837& 1.7737&  9.8669 &2.0536\\
 -3.7973& -1.0829& 3.4509& 1.8164&10.895& 2.0657\\
 -3.2585 &-1.0423&3.8181& 1.8512&11.311& 2.0699\\
 -2.7216& -0.99004&4.1853& 1.88&11.998& 2.0763\\
 -2.1875& -0.92043&4.5526& 1.9042&$+\infty$&2.182\\
\end{tabular}
\end{ruledtabular}
\end{table}

Now that the renormalization function has been determined, other
observables can be tested for the two particle system.
Interestingly, there is a numerically exact agreement between the
the external trap potential energy expectation values measured with
the exact wavefunction and with the mean-field renormalized
wavefunction. However, the one-particle density profiles calculated
using the exact wavefunctions and the mean-field renormalized
wavefunction are only in qualitative agreement, e.g. for
scattering lengths of large magnitude, where $|\overline{k}_f a_0|>>1$.

\section{Application to many-particle systems}

\begin{figure}[h]
\includegraphics[scale=0.6]{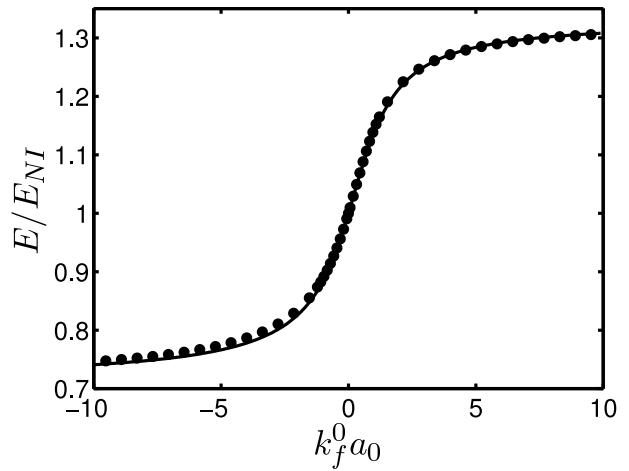}
\caption{Ratio of the total energy to the non-interacting energy,
for a spherically-trapped two-component degenerate Fermi gas in the
large N limit. The circles correspond to HF calculations for 2280
particles using $\zeta(k_f a_0)$, while the solid line corresponds
to either the variational solution, eq. (\ref{Evar}), or the TF
solution, eq. (\ref{Etot}), (the curves are indistinguishable on the
scale of the figure) using the approximate renormalization function
$\zeta_0(k_f a_0)$.} \label{EnergyTrap}
\end{figure}

This section presents different many-particle approximations for
which the renormalized scattering length can be used. The
renormalization procedure is designed to be used in the Hartree-Fock
approximation, however we will see that simpler approximations like
Thomas-Fermi (TF) or a variational trial wavefunction will yield
equally effective results in the large $N$ limit. The variational
wavefunction we will use is the noninteracting wavefunction rescaled
in the radial direction by a factor $\lambda$ as in
Eq.(\ref{psivariational}), where  $\lambda$ is the variational
parameter. As an example, we can see in Figure \ref{EnergyTrap} a
comparison for the ground state energy of a two-component Fermi gas
in an spherical trap in the large N limit. The result obtained with
the approximate $\zeta_0(k_fa_0)$ in conjunction either with a
variational trial wavefunction or else with the TF method are in
good agreement with the full HF calculation with the exact
$\zeta(k_fa_0)$. The difference between the results is mainly used
by the replace of the exact $\zeta(k_fa_0)$ by the approximate
$\zeta_0(k_fa_0)$. If we use the exact $\zeta(k_fa_0)$ for all the
methods, the energies agree in at least 3 digits. In systems having
a small number of particles, the HF method is, of course, most
reliable.

\subsection{Variational}

The simplest approximation \cite{rittenhouse2006hva} utilizes a
trial wavefunction that is a simple radial rescaling of the
noninteracting wavefunction:
\begin{equation}
   \label{psivariational}
  \Psi_{\lambda}(\mathbf{r}_1,\mathbf{r}_2,...,\mathbf{r}_N)= \frac{1}{\lambda^{3N/2}} \Psi_{NI}(\mathbf{r}_1/\lambda,\mathbf{r}_2/\lambda,...,\mathbf{r}_N/\lambda)
\end{equation}
The expectation value of the renormalized Hamiltonian (\ref{Ham})
can be separated into two terms,
$E(\lambda)=E_{HO}(\lambda)+E_{int}(\lambda,a_0)$, where
\begin{eqnarray}
  \label{Evariational}
 E_{HO}(\lambda)= \bra{\Psi_{\lambda}}\sum_i \left( -\frac{\hbar^2}{2m}\nabla^2_i+\frac{1}{2}m\omega^2\mathbf{r}^2_i\right)\ket{\Psi_{\lambda}}\nonumber\\
E_{int}(\lambda,a_0)=
\bra{\Psi_{\lambda}}\sum_{i<i'}\frac{4\pi\hbar^2a_{eff}}{m}
\delta(\mathbf{r}_i-\mathbf{r}_{i'})\ket{\Psi_{\lambda}}.
\end{eqnarray}

The energy of this trial wavefunction is calculated as a function of
the variational scale parameter $\lambda$ for the renormalized
Hamiltonian (\ref{Ham}). The non-interacting wavefunction is a
Slater determinant formed with the occupied spin-orbitals. The
$E_{HO}$ is simple to calculate, as it requires only a change of
variables to determine the $\lambda$-dependence in Eq.
\ref{Evariational} in conjunction with the known results of the
non-interacting ground state.
\begin{eqnarray}
  \label{EHO}
 E_{HO}(\lambda)= E_{NI}\left(\frac{1}{2 \lambda^2}+\frac{\lambda^2}{2}\right)
\end{eqnarray}
The interaction energy $E_{int}$ can be written in the following
form, when we apply the renormalization locally as a function of the
density:
\begin{equation}
   \label{Eintvar}
  E_{int}(\lambda,a_0)= \frac{4\pi\hbar^2}{m} \int
  \frac{\zeta(k_f^\lambda(\mathbf{r}) a_0)}{k_f^\lambda(\mathbf{r})}
  \rho^2_\lambda(\mathbf{r}) d\mathbf{r}
\end{equation}
In this equation $\rho$ is the density of one spin component and
$k_f^\lambda(\mathbf{r})\equiv(6\pi^2\rho_\lambda(\mathbf{r}))^{1/3}$.
In the large $N$ limit, the density of the non-interacting
wavefunction can be replaced by the TF density of the noninteracting
system \cite{Rittenhouse06}. The density corresponding to our trial
wavefunction is a simple radial rescaling, whereby the density in
the high-N limit is:
\begin{equation}
 \label{rholambda}
  \rho_{\lambda}(\mathbf{r})=\left\{ \begin{array}{lll}
  \frac{\sqrt{6 N}}{3 \pi^2 a_{ho}^3
  \lambda^3}\left(1-\frac{r^2}{2 a_{ho}^2 \lambda^2 (3
  N)^{1/3}}\right)^{3/2}&, & \mbox{if $r^2<R^2_c$} \\
   0&, & \mbox{otherwise}
  \end{array}\right.
\end{equation}
Here $N$ is the total number of particles and $R_c=\sqrt{2} a_{ho}
\lambda (3N)^{1/6}$ is the radius of the Fermi gas. In the large $N$
limit, the total energy can be expressed in units of the
noninteracting energy,
\begin{equation}
  \label{Evar}
 E/E_{NI}= \frac{1}{2 \lambda^2}+\frac{\lambda^2}{2}+ \frac{1}{\lambda^2}F\left(\frac{k_f^0 a_0}{\lambda}\right).
 \end{equation}
Here $\lambda$ is the scaling parameter, $k_f^0=\sqrt{2}(3
N)^{1/6}/a_{ho}$ is the Fermi momentum of the non-interacting system
at the trap center, and $F$ is
\begin{equation}
  \label{Ffun}
F(\gamma)=\frac{4^4}{9 \pi^2}\int_0^1 (1-x^2)^{5/2} x^2
\zeta\left(\gamma\sqrt{1-x^2}\right) dx.
 \end{equation}
This function must be calculated numerically unless further
approximations are made. The energy results obtained using Eq.
 (\ref{Evar}) are shown in Figure \ref{EnergyTrap}. In the unitarity
limit, the behavior can be calculated exactly:
\begin{equation}
  \label{Finf}
F(\gamma\rightarrow-\infty)=\frac{4^4 \zeta^{min}}{9 \pi^2}\int_0^1
(1-x^2)^{5/2} x^2 dx=\frac{5 \zeta^{min} }{9 \pi}.
 \end{equation}
\begin{figure}[h]
\includegraphics[scale=0.6]{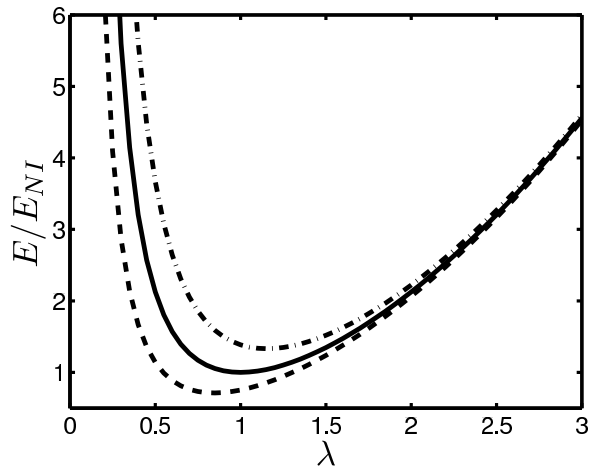}
\caption{Total energy as a function of $\lambda$, in units of the
noninteracting total energy. The solid curve corresponds to
$k_f^0a_0=0$, the dashed curve corresponds to $k_f^0a_0=-\infty$,
and the dotted-dashed curve corresponds to $k_f^0a_0=\infty$. The
minimum of the energy functional for $k_f^0a_0=-\infty$ occurs at
$\lambda=0.844$, which represents the ratio between the cloud radius
 at unitarity and the noninteracting cloud radius.} \label{VarEcurves}
\end{figure}
Three
curves predicted by Eq. (\ref{Evar}) are shown in Fig.
\ref{VarEcurves}. For the entire range of interactions, the energy
of the system (the minimum of the curve) remains finite, ranging
from $0.713 E_{NI}$ to $1.33 E_{NI}$, This shows how our
renormalization circumvents the collapse that would occur for the
bare Fermi pseudopotential.

\subsection{Thomas-Fermi results}

In this subsection we will review the TF approximation, using the
renormalization function. The TF approximation has been used to
study a 2-component Fermi gas with zero-range pseudo-potentials
\cite{roth2001}, but no renormalization has been considered.

Thomas-Fermi is of course a local density approximation. At each
position inside the trap, the wavefunction is approximated by a
Slater determinant of a set of plane wave orbitals, i.e., the
orbitals are characterized by four quantum numbers, the vector
momentum $\mathbf{k}$ and the spin. The orbitals are filled
uniformly up to a level $k_f(r)$ which is the same for spin up and
spin down fermions. The value of $k_f(r)$ will depend on the
distance $r$ from the trap center, and on the number of particles in
the system. For a uniform system, the value of $k_f$ is a constant
that characterizes the density of the system.

To calculate the local energy we need to sum over all the states at
that position.
 For example, the kinetic energy term $K$ for one-spin component is:
\begin{eqnarray}
  \label{pterm}
K=\frac{1}{2m}\sum_\mathbf{k} \braket{ \mathbf{k}
|\mathbf{p}^2|\mathbf{k}}=
\frac{\hbar^2}{2m}\sum_\mathbf{k} \mathbf{k}^2\braket{\mathbf{k}|\mathbf{k}}\nonumber \\
=\frac{V \hbar^2}{2m(2\pi)^3}\int k^2 d^3k =\frac{V \hbar^2 }{20
\pi^2m}k_f^5
\end{eqnarray}
Here $V$ is the volume of integration which will disappear when we
consider the local energy. This volume is small in comparison with
the external potential (in this case the trap) characteristic length
but is big enough to contain many particle. So, $k_f$ and $V_{ext}$
can be considered constant during the integration. The calculation
of the expectation value of an external trapping potential is then
straight forward and we obtain
\begin{equation}
  \label{Vterm}
\braket{V_{ext}}=\frac{V}{6\pi^2}V_{ext} k_f^3,
\end{equation}
which is just the number of particles times the external potential
at that position. In the case of the interaction term, the two
different spin components can be calculated by considering the
particles as indistinguishable, and calculating the direct and
exchange terms. But since the interaction is a delta function, we
can consider the particles as distinguishable with interactions only
between different species and obtain the same result. This latter
procedure is easier and we only need to consider the direct term.
\begin{multline}
  \label{Vintterm}
\braket{V_{int}}=\sum_{\mathbf{k}\mathbf{k}'} \braket{
\mathbf{k}\mathbf{k}' |V_{int}|\mathbf{k}\mathbf{k}'}\\
=\frac{1}{(2\pi)^6} \frac{4\pi a_{eff}\hbar^2}{m} \int_0^{k_f} \int_0^{k_f} \int \int \delta(\mathbf{x}-\mathbf{x}') d\mathbf{x}'d\mathbf{x} d^3k' d^3k  \\
=V \frac{4\pi a_{eff}\hbar^2}{m} \frac{k_f^3}{6\pi^2}
\frac{(k_f)^3}{6\pi^2} =V \frac{4\pi \hbar^2 }{m}
\frac{k_f^5\zeta(k_f a_0) }{(6\pi^2)^2}
\end{multline}

\begin{figure}[h]
\includegraphics[scale=0.6]{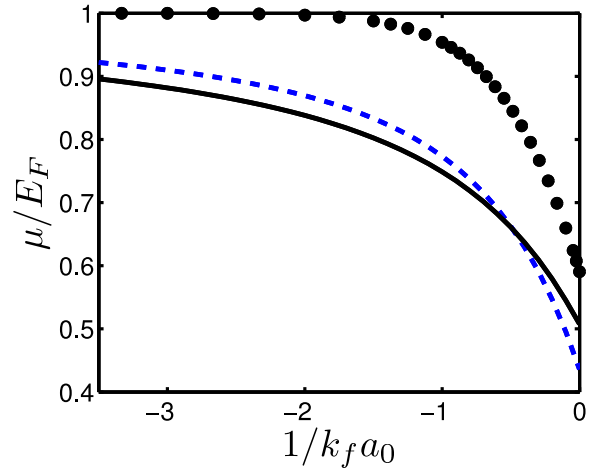}
\caption{(Color Online) Chemical potential in units of the Fermi
energy. The black solid line is the prediction obtained with the
renormalization function. The black circles represent the BCS
prediction and the blue dashed curve is the prediction obtained in
Ref. \cite{chin2005smf} }. \label{mufig}
\end{figure}

For the case of two equally-numerous spin components the local
energy (per unit volume) is
\begin{equation}
  \label{Elocal}
\mathcal{E}(k_f)=E(k_f)/V=\frac{\hbar^2}{2m}\frac{k_f^5}{5
\pi^2}+V_{ext}\frac{k_f^3}{3 \pi^2}+\frac{4\pi
\hbar^2}{m}\frac{k_f^5\zeta(k_f a_0) }{(6 \pi^2)^2}.
\end{equation}
In an infinite uniform system, where $V_{ext}=0$, the energy is:
\begin{equation}
  \label{Eunif}
\mathcal{E}(k_f)=\frac{\hbar^2}{2m}\frac{k_f^5}{5 \pi^2}+\frac{4\pi
\hbar^2}{m}\frac{k_f^5 \zeta(k_f a_0) }{36 \pi^4}.
\end{equation}
The ratio between the total energy and the non-interacting energy
has a simple form and only depends on $k_fa_0$,
\begin{equation}
  \label{Eunif2}
\mathcal{E}(k_f)/\mathcal{E}_{NI}(k_f)=1+\frac{10 \zeta(k_f a_0) }{9
\pi}.
\end{equation}
Using Eq. \ref{Elocal}, we can construct our energy functional by
integrating the local energy over all space.
\begin{multline}
  \label{Etot}
E=\int d\mathbf{r}\left(\frac{\hbar^2}{2m}\frac{k_f(\mathbf{r})^5}{5
\pi^2}+ V_{ext}(\mathbf{r})\frac{k_f(\mathbf{r})^3}{3 \pi^2}\right.\\
\left.+\frac{4\pi\hbar^2 }{m}\frac{k_f(\mathbf{r})^5\zeta(k_f a_0)
}{36 \pi^4}\right)
\end{multline}
To find the ground state we have to minimize the energy under the
constraint that the number of particles is fixed. This constraint
can be implemented by introducing a Lagrange multiplier $\mu_0$,
usually called the chemical potential. So, the minimization of Eq.
(\ref{Etot}) for fixed number of particles is reduced to the
minimization of
\begin{equation}
  \label{Lam}
\Lambda\equiv E-\mu_0 N=E-\mu_0 \int
d\mathbf{r}\frac{k_f(\mathbf{r})^3}{3 \pi^2},
\end{equation}
where variational parameter is $k_f(\mathbf{r})$. The necessary but
not sufficient condition for $k_f(\mathbf{r})$ to minimize $\Lambda$
is that
\begin{equation}
  \label{dLam}
\frac{\partial \Lambda}{\partial k_f(\mathbf{r})}=0.
\end{equation}
This condition leads to a relationship between the local chemical
potential, defined as
$\mu(\mathbf{r})\equiv\mu_0-V_{ext}(\mathbf{r})$, and the local
Fermi momentum $k_f(\mathbf{r})$,
\begin{equation}
  \label{mu}
\mu(\mathbf{r})=\frac{\hbar^2
k_f^2(\mathbf{r})}{2m}\left(1+\frac{10}{9\pi} \zeta(k_f(\mathbf{r})
a_0)+\frac{k_f(\mathbf{r}) a_0}{18\pi^2}
\zeta^\prime(k_f(\mathbf{r}) a_0)\right).
\end{equation}
The value of $\mu_0$ fixes the number of particles and, with this
relationship, we can calculate the density profile and the energy of
the system. Figure \ref{mufig} shows the chemical potential
dependence on $k_f a_0$ obtained with the renormalization function
and with other models. In Figure \ref{EnergyTrap} we can see the
energy obtained using eqs. (\ref{Etot},\ref{mu}) in the large N
limit.

In the unitarity limit, $a_0 \rightarrow -\infty$, we obtain
\begin{equation}
  \label{muunit}
\mu=\frac{\hbar^2 k_f^2}{2m}\left(1+\frac{10}{9\pi}
\zeta^{min}\right)
\end{equation}
  At unitarity, when the scattering length is much larger
than the inter-particle distance, the only relevant parameter is the
density \cite{heiselberg2001fsl,ohara2002osi}. Dimensional
analysis suggests that $\mu\propto \rho^{2/3}\propto k_f^{2}$. The
expected relation between $\mu$ and $k_f$ is usually written as
\begin{equation}
  \label{mubeta}
\mu=\frac{\hbar^2 k_f^2}{2m}(1+\beta).
\end{equation}
From our calculations this relations appears naturally with a
coefficient $\beta$, which is an universal parameter, of $\beta=10
\zeta^{min}/9\pi=-0.492$. This parameter $\beta$ has been studied
from many different perspectives. Table \ref{betavalues} shows
different experimental and theoretical values of $\beta$.
Interestingly, our $\beta$ value is consistent with most
experiments. To measure $\beta$ experimentally, Hulet and Thomas
groups measure the size of the cloud and compares it with the
noninteracting cloud. For, example, the result obtained by Hulet
group is $R_{U}/R_{NI}=0.825\pm 0.02$ which compares well with ours
$R_{U}/R_{NI}=0.844$ (Fig. \ref{VarEcurves}), obtained using both
variational or TF calculations we obtain.

\begin{table}
\caption{\label{betavalues} Experimental and theoretical predictions
of $\beta$.}
\begin{ruledtabular}
\begin{tabular}{ccccc}
 & &$\beta$& \\
\hline Experiments& $-0.54(5)$\footnotemark[1]  & $-0.64(15)$\footnotemark[2] & $-0.49(4)$\footnotemark[4] \\
Experiments& $-0.68^{+0.13}_{-0.10}$\footnotemark[3] & $-0.54^{+0.05}_{-0.12}$\footnotemark[5] \\
 QMC &  $-0.58(1)$\footnotemark[6] &  $-0.56(1)$\footnotemark[7]   \\
 Pad\'e asymptotes & -0.674\footnotemark[8] & -0.432\footnotemark[9]       \\
 Green's function & -0.545\footnotemark[10]& -0.599\footnotemark[11]   \\
 BCS& -0.41\footnotemark[12]    \\
 Other methods& -0.3\footnotemark[13] & -0.564\footnotemark[14] & -0.492\footnotemark[15]   \\
\end{tabular}
\end{ruledtabular}
\footnotetext[1]{Ref. \cite{partridge2006pap}} \footnotetext[2]{Ref.
\cite{bourdel2004esb}} \footnotetext[3]{Ref.
\cite{bartenstein2004cmb}} \footnotetext[4]{Ref.
\cite{kinast2005hcs}}
 \footnotetext[5]{Ref.\cite{Stewart06}}
\footnotetext[6]{Ref. \cite{astrakharchik2004esf}}
\footnotetext[7]{Ref. \cite{chang70qmc}}
\footnotetext[8]{Refs.\cite{bakerjr1999nmm,heiselberg2001fsl}}
\footnotetext[9]{Ref.\cite{bakerjr1999nmm}}
\footnotetext[10]{Ref.\cite{perali2004qcb}}
\footnotetext[11]{Ref.\cite{hu2006ess}}
 \footnotetext[12]{This is a
well known result, see for example Ref.\cite{perali2004qcb}}
\footnotetext[13]{Ref.\cite{bruun2004}}
\footnotetext[14]{Ref.\cite{chin2005smf}} \footnotetext[15]{Result
obtained in this paper with the renormalization function.}
\end{table}

\begin{figure}[h]
\includegraphics[scale=0.6]{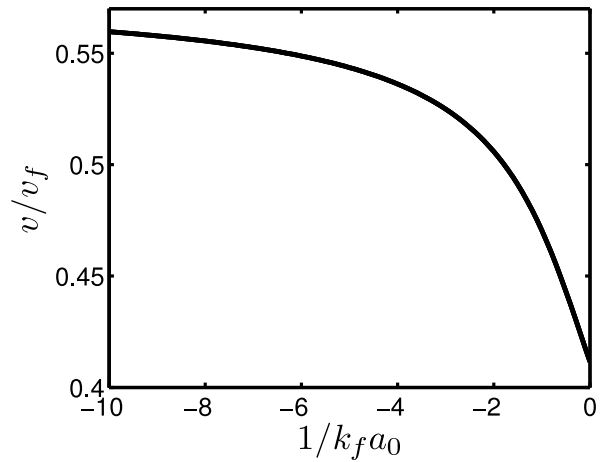}
\caption{The speed of sound is shown in units of
the Fermi velocity $v_f=\hbar
k_f/m$ for an uniform two-component Fermi gas,
as a function of $k_f a_0$.} \label{speedofsound}
\end{figure}

It is well established
\cite{houbiers1997ssa,ohara2002osi,zwierlein2005vas} that an
ultracold two-component Fermi system exhibits superfluidity. Even
though our renormalization scheme does not explicitly consider
superfluidity, it reproduces a number of properties of the Fermi gas
sensibly, including the equation of state and the chemical
potential. Consequently, these results can be used in the
hydrodynamic theory to extract information about dynamics of the
system, like the speed of sound or normal modes of excitation. For
example, the speed of sound in a uniform two-component system is
given by \cite{heiselberg2001fsl}
\begin{equation}
  \label{mubeta}
v^2=\frac{\hbar}{m}\frac{\partial}{\partial
\rho}\left(\rho^2\frac{\partial E/N}{\partial\rho}\right).
\end{equation}
Using Eq. (\ref{Eunif2}) we can thus evaluate the speed of sound,
which generates the results shown in Figure \ref{speedofsound}. The speed
of sound results reproduce the expected limiting behaviors. In the
noninteracting limit $v=v_f/\sqrt{3}$, while at unitarity $v=v_f
\sqrt{(1+\beta)/3}$ \cite{heiselberg2006smb}. This is one
example of a nontrivial observable quantity for this system
that can be predicted by this renormalization technique.
A comprehensive study of other observables
based on this approach will
be left for future publications.

\subsection{Hartree-Fock method}

The HF method for a many-particle system is an extension of the two
particle calculation done in Section II. The variational parameters
are the orbitals and the energy functional is:
\begin{multline}
  \label{Efunmp}
   \mathcal{E}(\psi)=\int \left(2\sum_{i}^{N/2}\psi_i(\mathbf{r})\left(
   -\frac{\hbar^2}{2m}\nabla^2+\frac{1}{2}m\omega^2\mathbf{r}^2\right)\psi_i(\mathbf{r})\right.\\
\left.+\frac{4\pi\hbar^2 k_fa_{eff}(k_f(\mathbf{r})
a_0)}{k_f(\mathbf{r})m} \rho(\mathbf{r})^2\right) d\mathbf{r}.
\end{multline}
In this approximation, the renormalization is done locally,
$k_f(\mathbf{r})= (6 \pi^2 \rho(\mathbf{r}))^{1/3}$. The
minimization procedure is lengthly and straight forward, so it will
not be presented here. By minimizing with respect to the set of
orbitals $\psi_i$, we obtain a set of nonlinear HF equations. These
are solved self-consistently. Figure \ref{EnergyTrap} shows results
for the HF energy of 2280 particles. This approximation is
particulary useful for systems with small number of particles, for
which the TF approximation has limited applicability. In Sec. IV
below, this method is used to obtain the energies of 8 fermions in a
trap.

\section{Results}

To compare the predictions based on our renormalized scattering
length with other methods we have carried out fixed node diffusion
Monte Carlo (FNDMC) simulations for equal mixture of different-spin
fermions. Interactions are considered only between different-spin
fermions, which are treated here as distinguishable particles. The
interaction potential is a purely attractive gaussian and its width
$d$ is chosen so that $\rho d^3\approx 10^{-4}$. The method follows
closely the approach of \cite{astrakharchik2004esf}.

The ground state wavefunction of two particles in a trap can be
separated exactly into a Jastrow term and non-interacting orbitals
\begin{equation}
  \label{2jas}
  \Psi_{p}(\mathbf{r_1},\mathbf{r_2})=
  \psi(\mathbf{r_1})\psi(\mathbf{r_2})J({\mathbf{r_1}-\mathbf{r_{2}}}),
\end{equation}
where $\psi$ is the non-interacting ground state orbitals. We will
show that this relation is valid when the center of mass
wavefunction  and the noninteracting orbitals are gaussians, like in
the trapped. For this discussion $\hbar=m=\omega=1$ and we will not
consider normalization factors. If we define $\mathbf{R}_{CM}=(1/2)
(\mathbf{r_1}+\mathbf{r_{2}})$ and $\mathbf{r}=
(\mathbf{r_1}-\mathbf{r_{2}})$ we know that the pair wave function
can be separated in center of mass and relative coordinate term, $
\Psi_{p}(\mathbf{r_1},\mathbf{r_2})=\Psi(\mathbf{R}_{CM})\phi(\mathbf{r})$.
Also, we know that the center of mass wavefunction is unaffected by
the two-body interaction and is
$\Psi(\mathbf{R}_{CM})=\exp(-\mathbf{R}_{CM}^2)=\exp(-\mathbf{r}_1^2/2)
\exp(-\mathbf{r}_2^2/2)\exp(\mathbf{r}^2/4)=\psi(\mathbf{r}_1)\psi(\mathbf{r}_2)\exp(\mathbf{r}^2/4)$,
so the total wavefunction is $\Psi(\mathbf{R}_{CM})
\phi(\mathbf{r})=\psi(\mathbf{r}_1)\psi(\mathbf{r}_2)
\exp(\mathbf{r}^2/4)\phi(\mathbf{r})=\psi(\mathbf{r}_1)\psi(\mathbf{r}_2)J(\mathbf{r})$,
where $J(\mathbf{r})=\exp(\mathbf{r}^2/4)\phi(\mathbf{r})$. The
evaluation of the relative coordinate wavefunction,
$\phi(\mathbf{r})$, requires in general of a numerical calculation.

We use this Jastrow term to construct the many-body wavefunction.
This wavefunction is usually called Jastrow-Slater wavefunction.
Here, $i$ and $i'$ correspond to different-spin fermions. The
non-interacting wavefunction is a Slater determinant formed with the
harmonic oscillator orbitals.
\begin{equation}
  \label{jaswf}
  \Psi(\mathbf{r}_1,\mathbf{r}_2,...,\mathbf{r}_N)=
  \prod_{ii'}J({\mathbf{r_i}-\mathbf{r_{i'}}}) \Psi_{NI}(\mathbf{r}_1,
  \mathbf{r}_2,...,\mathbf{r}_N)
\end{equation}
We have also used a BCS-type many-body wavefunction constructed with
pair wavefunctions for FNDMC
 \begin{multline}
  \label{pairtypewf}
  \Psi(\mathbf{r}_1,\mathbf{r}_1',...,\mathbf{r}_{N/2'})=\\
 \mathcal{A}\left\{\Psi_p(\mathbf{r}_1,\mathbf{r}_{1'})\Psi_p(\mathbf{r}_2,\mathbf{r}_{2'})...\Psi_p(\mathbf{r}_{N/2},\mathbf{r}_{N/2'})\right\},
\end{multline}
where $\mathcal{A}$ is the antisymmetrizer operator. This trial
wavefunction leads to good results on the BEC limit but in the BCS
regime the Jastrow-Slater wavefunction produces lower energies.
\begin{figure}[h]
\includegraphics[scale=0.6]{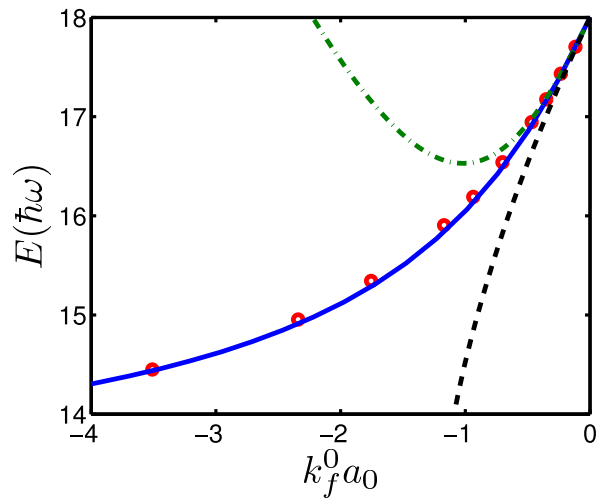}
\caption{(Color Online) The total energy of 8 fermions in a trap is
shown in oscillator units as a function of $k^0_fa_0$. FNDMC results
are shown in open red circles while full blue line corresponds to HF
results using the effective renormalized scattering length. The
dashed and the dash-dotted curves correspond respectively to
solution use first-order, or first- and second-order terms, in an
expansion into powers of $k_fa_0$. See the discussion of Eqs.
(\ref{ExpFer},\ref{Efunexp}).} \label{p8res}
\end{figure}

We have calculated FNDMC energies for 8 particles in the BCS side of
the crossover. In Fig. \ref{p8res} these energies are compared with
HF calculations including the first- and second-order corrections in
the $k_fa_0$ expansion \cite{HuangYang}, and then also with full HF
calculations using the renormalized scattering length directly.
\begin{equation}
   \label{ExpFer}
  E_{int}/N=\frac{\hbar^2 k_f^2}{m}\left(\frac{k_f a_0}{3\pi}+
  \frac{6(11-2\ln2)}{105\pi^2}(k_f a_0)^2+...\right)
\end{equation}
The idea of using this type of expansion to construct energy
functionals has been applied for bosons \cite{blume2001qcg,
braaten1997qcg}. The expansion (\ref{ExpFer}) can be introduced
locally in variational treatments, which yields an energy
functional,
\begin{multline}
  \label{Efunexp}
   \mathcal{E}(\psi)=\int \left(2\sum_{i}^{N/2}\psi_i(\mathbf{r})\left( -\frac{\hbar^2}{2m}\nabla^2+\frac{1}{2}m\omega^2\mathbf{r}^2\right)\psi_i(\mathbf{r})\right. \\
 \left.+\frac{4\pi\hbar^2a_{0}}{m} \rho(\mathbf{r})^2+a_0^2\frac{12(11-2\log(2))}{105\pi^2}(6\pi^2)^{4/3}\rho(\mathbf{r})^{7/3}\right)d\mathbf{r}.
\end{multline}
where $\rho=\rho_{\uparrow}=\rho_{\downarrow}$. If we only consider
the first term in Eq. \ref{ExpFer} we obtain the Fermi
pseudo-potential contribution.

To study the weak interacting limit, previous authors
\cite{bruun1998ifg,roth2001} have considered the Fermi
pseudopotential approximation, which is only the first term in the
energy expansion (\ref{ExpFer}). Applying the expansion
(\ref{ExpFer}) in the local density approximation is a convenient
way to introduce higher order corrections to mean field theories. We
can obtain an expansion of the density-dependent renormalization
function using Eq. (\ref{ExpFer}), in this case
\begin{equation}
   \label{kfaexp}
  \zeta(k_fa_0)=k_f a_0+\frac{6(11-2\ln2)}{35\pi}(k_f a_0)^2+...
\end{equation}
Insertion of this result into Eq. (\ref{Ham}), with the local
density approximation and a Slater determinant wavefunction, gives
Eq. (\ref{Efunexp}).

A power expansion of $\zeta(k_fa_0)$ obtained by the renormalization
method should agree with this expansion. The first term is
reproduced exactly but the second one is only in qualitative
agreement. While the coefficient of the second order expansion in
Eq. (\ref{kfaexp}) is approximately 0.525, in the density
renormalization from Section II the coefficient is 0.422. This
disagreement may be due to the level of approximation of the density
renormalization procedure.

We find very good agreement between the mean-field results
calculated using the renormalized interaction developed in this
paper, and the FNDMC (Fig. \ref{p8res}). The variational methods
including the perturbative corrections (Eq. \ref{Efunexp})  show
good agreement in the small $k_fa_0$ region, deviating from the
FNDNC results when the corrections to the expansion (\ref{ExpFer})
become important.

\begin{figure}[h]
\includegraphics[scale=0.6]{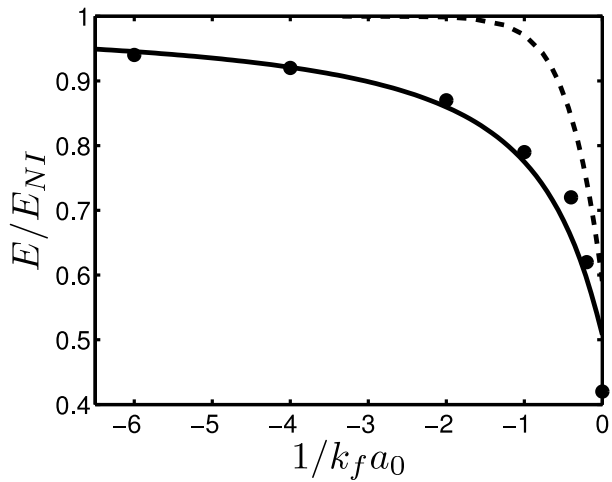}
\caption{$E/E_{NI}$ for an infinite homogeneous Fermi gas $k_fa_0$
in the mean-field approach (solid line). The dashed curve
corresponds to the local density BCS solution and the circles
correspond to FNDMC results obtained in
\cite{astrakharchik2004esf}.} \label{Eperpar}
\end{figure}

It is also possible to make comparisons with other quantum Monte
Carlo calculations. Astrakharchik and coworkers
\cite{astrakharchik2004esf}  have studied a homogenous doubly
degenerate Fermi gas using FNDMC methods. In their calculations,
they considered up to 60 particles. We compare the energy of this
system as obtained using the density renormalization procedure in
Eq. (\ref{Eunif2}). A comparison between the two calculations and
local density BCS result \cite{leggett1980mtt,marini1998ebs} is
shown in Fig. (\ref{Eperpar}).

\section{Conclusions}


It is the goal of most many-body theoretical studies to derive
predictive power for numerous observables of interest, using simpler
methods that bypass the actual calculation of this "true" ground
state wavefunction for the trapped atomic gas.  At the heart of many
such treatments are the following two steps: {\it (i)} replacement
of the two-body potential energy by a zero-range Fermi
pseudopotential, followed by {\it (ii)} a mean-field wavefunction
ansatz and the computation of observables.  The basic level of
description for a Bose gas incorporates no correlations whatsoever.
For a Fermi gas, correlations are generally treated at either the
bare minimalist level of exchange correlations alone, using a single
Slater determinantal wavefunction.  A more sophisticated level is
often considered for a system of mutually attractive fermions, which
are frequently described with BCS-type correlations built into the
description.  One way of visualizing the value of a Fermi-type
zero-range pseudopotential adopted in most such theories is to
remember that it has been specifically designed to give a meaningful
interaction energy for each pair of particles even when the
wavefunction structure is too simplistic to incorporate any
appreciable correlations.

The present article presented an alternative implementation of this
general philosophy.  We developed a procedure for renormalizing the
coefficient of a zero-range potential, based entirely on an analysis
of the nonperturbative two-body system solved first with and then
without wavefunction correlations.  When we applied this procedure
to the many-body Fermi gas, it gives agreement with the standard
dilute gas limit, an important prerequisite for any realistic
theory.  But in addition, it is able to treat higher densities $n$,
including the regime $|n {a_0}^3| >> 1$. We studied a number of
observables that have been explored both experimentally and
theoretically in the BCS-BEC crossover regime, and found good
agreement using our renormalized Hartree-Fock approach all the way
to the unitarity limit $a_0 \rightarrow \infty$. Perhaps
surprisingly, this good agreement is achieved without incorporating
explicit BCS-type correlations into the many-body wavefunction.  One
result of this study is an approximate expression for the
renormalization function in closed analytical form that may prove to
be useful in other studies of the two-component degenerate Fermi
gas. Another interesting result is that at unitarity, the chemical
potential exhibits the expected density dependence characterized by
the parameter $\beta=-0.492$, which, interestingly, is consistent
with recent experiments
\cite{partridge2006pap,bourdel2004esb,kinast2005hcs,Stewart06}.

In order to study the complete BCS-BEC crossover, future
improvements of this theory should include a more flexible many-body
wavefunction which can represent a Fermi gas in the weak interacting
region and a gas of Bose molecules in the BEC region.

\begin{acknowledgments}
We thank S. Rittenhouse for access to unpublished results.
Discussions with D. Blume, L. Radzihovsky, V. Gurarie, Tin-Lun Ho
and J. Thomas are appreciated. This work was supported in part by
NSF.
\end{acknowledgments}

\end{document}